\documentclass[sigplan,nonacm]{acmart}

\usepackage{amsmath,amsfonts}
\usepackage{newtxtext,newtxmath}
\usepackage{graphicx}
\usepackage{textcomp}
\usepackage{enumitem} 
\usepackage{tikz}
\usepackage[normalem]{ulem}
\usepackage[scaled=0.825]{beramono}
\usepackage{listings}
\usepackage{listing}
\usepackage{xspace}
\usepackage{multirow}
\usepackage{adjustbox}
\usepackage{enumitem}
\usepackage{csquotes}
\usepackage{algorithm}
\usepackage{algpseudocode}

\usepackage{xcolor}

\lstdefinelanguage{gas}{
  sensitive=true,
  morekeywords={movzwq,cmp,je,inc,movw},
  morecomment=[l]{//},
}

\lstdefinestyle{gasstyle}{
  language=gas,
  basicstyle=\ttfamily\footnotesize,
  columns=fullflexible,
  keepspaces=true,
  showstringspaces=false,
  xleftmargin=17pt,
  numbers=left,
  numberstyle=\tiny,
  numbersep=8pt,
  breaklines=true,
  keywordstyle=\bfseries,
  commentstyle=\itshape,
  captionpos=b,
}
\microtypecontext{spacing=nonfrench}

\newcommand{\name}{Clove\xspace}
\newcommand{\names}{Clove's\xspace}
\newcommand{\para}[1]{\noindent\textbf{#1 }}

\newcommand*\circled[1]{\tikz[baseline=(char.base)]{
            \node[shape=circle,draw,inner sep=1pt] (char) {#1};}}

\newcommand{\bi}{\begin{itemize}}
\newcommand{\ei}{\end{itemize}}

\newcommand{\eat}[1]{}

\renewcommand\footnotetextcopyrightpermission[1]{}
\acmConference[]{}{}{}

\begin{document}\pagestyle{plain}

\title{
\name: Object-Level CXL Memory Management in Managed Runtimes
}

\author{Sam Son}
\affiliation{%
  \institution{UC Berkeley}
    \city{Berkeley}
  \country{USA}
} 
\author{Zhihong Luo}
\affiliation{%
  \institution{UC Berkeley}
  \city{Berkeley}
    \country{USA}
} 
\author{Wen Zhang}
\affiliation{%
  \institution{UC Berkeley}
  \city{Berkeley}
    \country{USA}
} 
\author{Sylvia Ratnasamy}
\affiliation{%
  \institution{UC Berkeley}
  \city{Berkeley}
    \country{USA}
} 
\author{Scott Shenker}
\affiliation{%
  \institution{UC Berkeley and ICSI}
  \city{Berkeley}
    \country{USA}
} 

\begin{abstract}
Object-level management of tiered memory has been studied to address the inefficiencies in page-based systems. However, object-level management for CXL-tiered memory remains underexplored due to CXL's tight performance budget and load/store interface. As a result, existing approaches remain limited in scope, primarily targeting unmanaged-language applications with bespoke runtimes or compiler support.

This paper identifies and explores a new design point for object-level CXL management: managed languages and their runtimes. The key observation is that existing managed runtimes already provide highly optimized mechanisms for problems closely related to object-level management, including object relocation and dynamic code generation. However, they still lack the features needed for tiered memory management, such as hotness tracking and relocation policies, and thus must be carefully extended to fully realize this direction.

We present \name, a system that extends existing managed runtimes to support object-level CXL management for managed-language applications. \name combines profile-guided object hotness tracking with object relocation techniques and policies. Our JVM prototype demonstrates that this extension enables high utilization of fast-tier memory while bounding runtime overhead, reducing application slowdown by 22--84\% compared to page-based systems.
\end{abstract}

\maketitle

\section{Introduction}
\label{sec:intro}

The longstanding effort to build tiered-memory systems has recently accelerated with the advent of Compute Express Link (CXL). By offering significantly higher performance than previous memory-expansion approaches~\cite{liu2024dissecting,samsung-cxl}, CXL has spurred substantial research interest and growing industry adoption~\cite{intelBreakingMemory,li2023pond}.
As with any tiered memory system, efficient use of CXL memory requires maximizing the \textit{fast-tier hit ratio}, i.e., the fraction of memory accesses served by the fast memory tier. Achieving this requires solving two key design problems: \textit{hotness tracking}, which monitors the recency and frequency of memory accesses to distinguish hot data from cold data, and \textit{data relocation}, which moves hot data to the fast tier and cold data to the slow tier at an appropriate granularity.

Existing CXL memory management systems typically rely on virtual memory and use pages as the unit of both hotness tracking and relocation (see, for example, \cite{lee2023memtis,raybuck2021hemem,vuppalapati2024tiered,xiang2024nomad,duraisamy2023towards,maruf2023tpp,li2023pond,zhong20242lm,autonuma,agarwal2017thermostat,maruf2022multi}) because doing so preserves \textit{application transparency} and incurs low runtime overhead, i.e., \textit{efficiency}. However, such techniques face the fundamental problem that data on a single page may have very different access frequencies, a phenomenon we call \textit{intrapage hotness skew}; thus, even an optimal placement of pages cannot achieve the optimal fast-tier hit ratio. This problem is exacerbated by the extensive use of huge pages~\cite{redhat52xA0HugePages,java-largepages} (\S\ref{subsec:intrapage}).

Object-level management has been studied as a solution to this problem in the broader context, but CXL's characteristics make it particularly challenging for two reasons. \textit{A higher bar for efficiency}: CXL is much closer to local memory in performance than prior slow tiers, so the overhead of tracking and relocating many objects can easily outweigh the benefit of improved placement. \textit{No anchor point}: CXL memory is accessed through ordinary loads and stores, leaving no natural software interception point for transparent management. We elaborate on these points in \S\ref{subsec:cxl}.

As a result, object-level CXL memory management remains underexplored. To our knowledge, only one concurrent work studies object-level CXL management~\cite{banakar2026obase}, and it follows an approach proposed by prior tiering systems: it asks developers of unmanaged-language applications to rewrite important data structures using custom APIs, then builds dedicated runtime or compiler support to track and relocate the corresponding objects. This approach can be efficient because it narrows the number of objects to track, and it creates an explicit anchor point for management; however, it is not transparent because it still requires application rewriting.

In this paper, we explore a new design point for object-level CXL memory management that is based on \textit{extending existing managed runtimes}, such as the Java Virtual Machine (JVM) and .NET Common Language Runtime (CLR). 
While prior work starts from unmanaged languages and adds new runtime or compiler support to recover object-level visibility and efficient control, we instead start from the opposite direction: managed runtimes already have object-level visibility and efficient control mechanisms, and we ask whether they can be extended to manage CXL tiered memory efficiently and transparently at object granularity.

This direction is important because it targets managed-language applications, a large class of production software that has received limited attention from prior object-level tiering systems. It is also promising because managed runtimes already solve problems closely related to object-level CXL management (\S\ref{subsec:opportunity}): modern garbage collectors provide optimized support for relocating objects for heap compaction, and JIT-compilation-based execution provides a way to observe runtime behavior and selectively instrument code based on it, which is essential for transparent hotness tracking. Together, these mechanisms offer a natural starting point for efficient and transparent object-level management.

However, managed runtimes do not provide object-level CXL management out of the box, posing the following technical challenges (\S\ref{sec:overview}). First, they still require efficient and transparent object hotness tracking for CXL. Existing tracking methods either require developer assistance or become too expensive when applied directly to objects. Second, the managed runtime lacks visibility into the physical tiers, so it must coordinate with the operating system to perform physical tier placement. Lastly, while object relocation mechanisms can be reused, how exactly they are repurposed matters. %

Motivated by these observations and challenges, we present \name, a transparent and efficient object-level CXL memory management system for \textit{managed-language applications}. \name addresses the challenges above by carefully extending existing runtime mechanisms (\S\ref{sec:overview}). First, it introduces \textit{profile-guided object hotness tracking}: \name uses hardware-assisted profiling to identify the small set of load instructions responsible for most LLC misses and then tracks only the object accesses triggered by those instructions. This approach captures only accesses relevant to CXL tiering, providing accurate and transparent object hotness tracking with low overhead. Second, \name uses a hybrid relocation approach that combines runtime-level \textit{hot-object compaction}, which compacts hot objects into contiguous virtual pages, with existing page-level systems, which migrate the corresponding physical pages between tiers. This separation keeps the runtime independent of physical page placement and, as an additional benefit, allows large objects that span multiple pages to be managed at page granularity. Third, \name introduces policies for hot-object compaction that carefully select objects to relocate, improving fast-tier utilization while bounding relocation overhead. We detail \names design in \S\ref{sec:design}.

We prototype \name on top of OpenJDK 21 (\S\ref{sec:impl}) and evaluate it using three memory-intensive Java applications under realistic workloads (\S\ref{sec:eval}). Our evaluation demonstrates that \name reduces application slowdowns by 22--84\% compared to state-of-the-art page-based systems without requiring any source-code changes. These improvements result from object-level management enabled by \name's key design ideas, while \name incurs little additional runtime overhead.

Although \name is prototyped in the JVM, the overall approach is not specific to Java. \name relies on runtime capabilities common to several managed runtime implementations: object-level memory management, moving garbage collection, and JIT-based code generation. Thus, we believe the same principles can apply to other managed runtimes with such capabilities, such as C\#/.NET and PyPy.

In summary, this paper makes the following contributions:
\begin{itemize}[leftmargin=8pt,itemsep=0pt,parsep=0pt,topsep=0pt,partopsep=0pt]
\item We identify extending existing managed-language runtimes as a new design point for efficient and transparent object-level CXL memory management;
\item We identify the challenges in extending managed runtimes to support object-level CXL management and address them through novel object hotness tracking, a hybrid page/object approach, and systematic hot-object compaction; and
\item We prototype these ideas in \name and show that \name transparently reduces application slowdowns by 22--84\% in managed-language applications compared to state-of-the-art page-based systems under intrapage hotness skew.
\end{itemize}

\section{Background and Motivation}
\label{sec:motiv}
In this section, we investigate the problem of intrapage hotness skew in page-based CXL systems and underscore the opportunities in extending existing managed runtimes for object-level CXL management.
\label{subsec:background-space}
\begin{figure}[t]
    \centering
    \includegraphics[width=0.95\columnwidth]{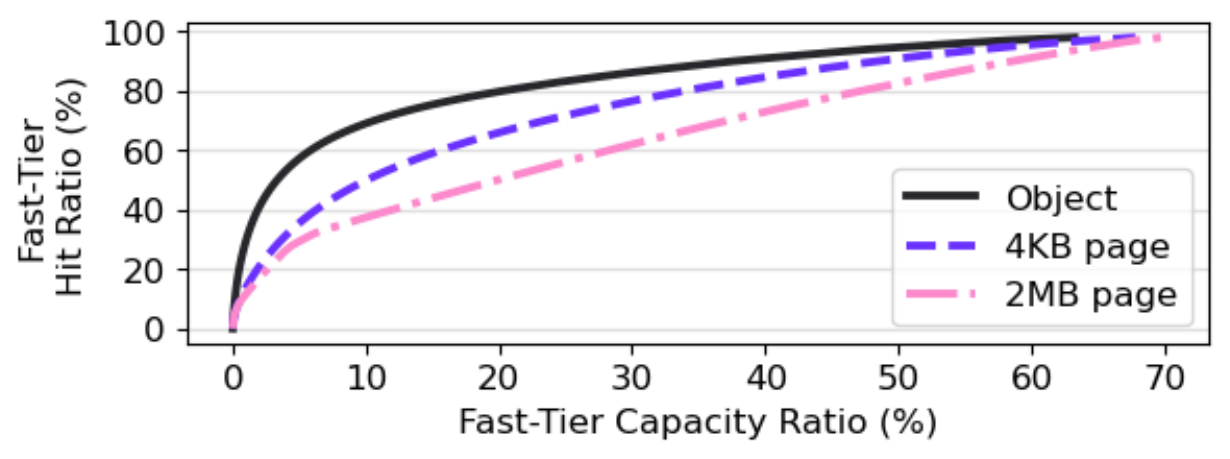}
    \vspace{-0.15in}
    \caption{Fast-tier hit ratio under oracle placement with objects (256 B), 4 KB pages, and 2 MB pages as relocation units. Setup: a key-value cache with a Zipfian distribution.}
    \vspace{-0.25in}
    \label{fig:space-waste}
\end{figure}
\subsection{CXL-based Tiered Memory}
\label{subsec:cxl}
\noindent Compute Express Link (CXL) is an interconnect that enables memory expansion by attaching additional memory devices outside the processor socket while maintaining a load/store memory interface. CXL-tiered memory systems aim to manage a memory hierarchy consisting of fast, locally attached memory and slower, out-of-socket memory connected via CXL. We refer to the former as \textit{local} or \textit{fast-tier memory} and the latter as \textit{CXL} or \textit{slow-tier memory} throughout the paper.

Two properties of CXL memory shape the design of these systems: its relatively small performance gap from local memory and its load/store access interface. CXL memory is only about 2--4$\times$ slower than locally attached DDR5 memory~\cite{li2023pond,zhong20242lm,sun2023demystifying,liu2024dissecting}, a much smaller gap than in network-attached memory or storage. In addition, CXL memory is accessed through ordinary load and store instructions, which provide no natural interception point for implementing software management mechanisms, unlike the swapping interface in storage and far-memory systems. These properties imply that CXL memory management must be highly efficient, since management overhead can easily negate the benefit of improved data placement, and that implementing fully transparent management mechanisms is challenging.

\subsection{Intrapage Hotness Skew in Page-level Systems}
\label{subsec:intrapage}
Existing CXL management systems address these challenges by building on virtual memory~\cite{lee2023memtis,vuppalapati2024tiered,xiang2024nomad,duraisamy2023towards,maruf2023tpp,li2023pond,zhong20242lm,autonuma,song2025hybridtier,ren2024mtm,xu2024flexmem}. They achieve transparency by operating at the OS level and provide efficient data relocation through page tables and hardware support such as the MMU and TLB~\cite{hennessy2011computer}. However, since common page sizes (4 KB and 2 MB) in modern architectures are much larger than most program objects, each page can contain many objects with different levels of hotness; we call this \textit{intrapage hotness skew}.

To quantify the inefficiency caused by this skew, we evaluate the fast-tier hit ratio achievable by an \textit{oracle} policy when data can be relocated at three different granularities: 4 KB pages, 2 MB pages, and objects (256 B). We use a key-value cache, Ehcache, with Zipfian key accesses. By an oracle policy, we mean the placement we would choose if we knew the future access pattern over the measurement period: it ranks all relocation units by their eventual access frequency and fills the fast tier with the hottest units first. To emulate this policy, we first collect a trace of memory accesses from a 5-minute run using PEBS samples of L3 cache misses. From this trace, we compute the access frequency of each relocation unit using the address in each sample. We then rank the units by access frequency and assume that the most frequently accessed units are placed in the fast tier until the fast-tier capacity is exhausted. Finally, we scan the trace again and compute the fast-tier hit ratio as the fraction of sampled accesses from the relocation units selected for fast-tier placement. We repeat this procedure while increasing the PEBS sampling rate until the resulting hit-ratio curves saturate.

Figure~\ref{fig:space-waste} shows that object-level management achieves 19.6 and 15.4 percentage points higher fast-tier hit ratios than 4 KB pages at fast-tier capacity ratios of 10\% and 20\%, respectively. This difference stems from the inefficient packing inherent to page-level migration; a single 4 KB page holds a dozen key-value pairs with varying memory access frequencies; thus, without careful object placement, pages placed in the fast tier inevitably include cold objects that result in lower hit ratios. The use of huge pages exacerbates the space waste. Each 2 MB page contains about 7000 key-value pairs, making most pages similarly warm according to their aggregate hotness. As a result, the hit ratio in the 2 MB curve increases almost linearly after 5\%, remaining much lower than the others. %

\subsection{Why Existing Runtimes Are a Good Starting Point for Object-Level Management}
\label{subsec:opportunity}
\para{Object-level Management} 
\emph{Object-level} management has been explored as a way to address intrapage hotness skew or related problems, e.g., write amplification~\cite{calciu2021rethinking}, in the broader tiered memory context~\cite{ruan2020aifm,wang2022memliner,wang2020semeru,amaro2020can,gu2017efficient,al2020effectively,zhou2022carbink,banakar2026obase}. A common approach in this line of work is to target unmanaged languages such as C/C++ and expose object-level control to developers through dedicated runtime or compiler support. One concurrent work applies a similar approach to CXL memory management~\cite{banakar2026obase} (see \S\ref{sec:rel} for details).

\para{Opportunities in Managed Runtimes} As we motivate in \S\ref{sec:intro}, despite this rich line of work, an important and natural direction remains unexplored: targeting managed-language applications and using existing managed runtimes as the starting point for object-level CXL management. %

In this section, we highlight the overlap between the problems shared by object-level CXL management and managed runtimes by elaborating on two such problems. First, managed runtimes already solve the problem of object relocation for garbage collection. The goal of garbage collection is to identify dead objects and reclaim free space. However, when live and dead objects are interleaved, simply freeing dead objects leaves the heap fragmented and limits how much contiguous space can be reused. Modern garbage collectors address this problem by relocating live objects into contiguous empty regions on the heap, thereby coalescing free space. 

Decades of research have made this object-relocation process efficient. For example, rather than accessing objects through indirect handles, which would add an extra memory access to every object access, modern managed runtimes such as the JVM access objects directly and pay relocation costs only when needed by updating references through heap scans or barriers~\cite{yang2022deep,oracleHotSpotVirtual}. This design is especially important for CXL, where access latency is close to local-memory latency.

Second, managed runtimes provide \textit{dynamic code generation} through \textit{just-in-time (JIT) compilation}. %
Aggressively optimizing all code upfront would delay startup and waste compilation effort on cold code, while interpreting all code would leave hot paths under-optimized. Modern managed runtimes address this tradeoff through tiered compilation: they initially execute methods using an interpreter, monitor execution to identify hot methods, and then JIT-compile only those methods with stronger optimizations.

This capability is essential for efficient and transparent object-level management. Prior object-level management systems show that selectivity is key to keeping tracking overhead low: instead of tracking every object access, they use developer assistance to focus on accesses that are likely to matter. Achieving this selectivity transparently requires observing program behavior during execution and modifying or generating code without developer intervention. The existing framework provides exactly this opportunity.

In summary, existing managed runtimes provide mature and performant implementations of key techniques required for object-level management. This makes them the natural starting point for adding support for CXL tiered memory.

\section{Our Proposal: \name}
\label{sec:overview}
While existing managed runtimes provide a natural foundation for object-level CXL management, turning them into a complete system still poses several technical challenges. In this section, we describe these challenges and present the core ideas that \name uses to address them. At a high level, \names design shows how to minimally extend the runtime to be "hotness-aware" and what the right division of labor is between the runtime and page-based systems.

\begin{figure}[t]
    \centering
    \includegraphics[width=0.95\columnwidth]{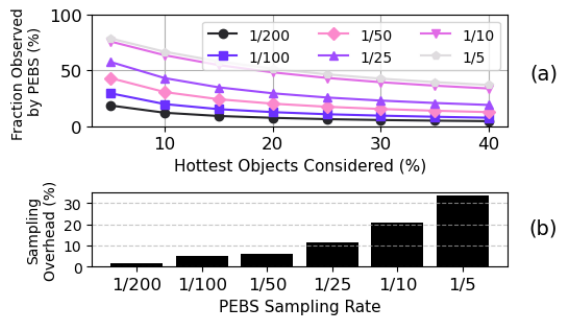}
    \caption{(a) Object observability with PEBS. For each hottest-object set on the x-axis, we report the fraction of objects in that set observed by PEBS during a 1-minute run. (b) Runtime overhead of PEBS with different sampling rates.}
    \vspace{-0.25in}
    \label{fig:pebs-overhead}
\end{figure}
\subsection{Object-level Hotness Tracking}
\para{\emph{Challenge}:} Although managed runtimes track object state to determine liveness, they do not provide object-level hotness tracking. The ideal hotness-tracking system must satisfy three requirements: \emph{tracking accuracy} to enable a high fast-tier hit ratio; \emph{efficiency}, incurring low runtime overhead in the CXL context; and \emph{transparency}. Existing approaches fall short of satisfying all of these requirements. 

First, as mentioned earlier, techniques based on developer assistance can narrow the range of objects to track, but they sacrifice transparency. Some prior work in far memory uses profiling to automate parts of this process~\cite{guo2023mira,tauro2024trackfm}; however, these systems rely on software interception points, i.e., swapping, that are not available for CXL memory accesses, and they still require offline profiling runs.

Another set of candidates comes from page-based systems: page faults~\cite{autonuma,maruf2023tpp,kim2021exploring}, page-table scans~\cite{yan2019nimble,duraisamy2023towards}, or hardware-assisted sampling~\cite{lee2023memtis,raybuck2021hemem,song2025hybridtier}. Since page faults and page-table scans expose only page-level accesses, we evaluate the applicability of hardware-assisted sampling, such as Intel PEBS~\cite{intelTimedProcess} and AMD IBS~\cite{drongowski2010incorporating}.

Hardware-assisted sampling is poorly suited to object-level hotness tracking: sampled addresses must be mapped back to objects, and the large number of objects requires extremely high sampling rates to achieve even moderate coverage, adding large CPU overhead~\cite{xiang2024nomad,lepers2023johnny}.
To demonstrate the latter, we use the same key-value cache setup as in \S\ref{subsec:intrapage} and collect L3 cache-miss samples using PEBS to measure what fraction of the hottest $N\%$ objects PEBS can observe with varying sampling rates. Figure~\ref{fig:pebs-overhead}(a) shows how well PEBS identifies a target set of hot objects at each sampling rate, while Figure~\ref{fig:pebs-overhead}(b) shows the corresponding runtime overhead. With a sampling rate of 1-in-10, PEBS incurs roughly 20\% runtime overhead but captures only 60\% of the hottest 10\% of objects.

\para{\emph{Our solution}:} To satisfy the three requirements, \name introduces \textit{profile-guided object hotness tracking}.
The key insight is that, although directly estimating object hotness with hardware sampling is ineffective due to the large number of objects, the number of instructions responsible for most main-memory accesses---namely, last-level cache misses---is much smaller. \name therefore uses hardware-assisted sampling not to track object hotness directly, but to identify load instructions responsible for most L3 cache misses. \name then uses JIT compilation to insert hotness-tracking logic only at these delinquent loads. While using PEBS to filter instructions is similar in spirit to prior profile-guided optimizations~\cite{luo2024harvesting,apt-get,chen2016autofdo}, \name applies this idea to object-level hotness tracking. This design achieves efficiency through selectivity and accuracy by continuing to observe accesses relevant to CXL tiering, all without breaking transparency (\S\ref{subsec:design-profiler}).

The efficiency of this design still hinges on keeping hotness tracking logic low overhead. To this end, \name repurposes unused bits in per-object metadata already maintained by managed runtimes, following prior work~\cite{wang2019panthera,akram2018write,nguyen2024polar,yang2020improving}. Updating this compact counter requires only a few simple instructions, and the object header is usually already in the L1 cache when the object is accessed. Using object headers also avoids separate metadata storage. \name further applies additional optimizations to reduce tracking overhead, achieving near-zero runtime overhead for hotness tracking (\S\ref{subsec:design-hotness-tracking}).

\subsection{Object Relocation Mechanisms}
\para{\emph{Challenge}:} Managed runtimes provide an efficient and transparent mechanism for relocating objects for GC purposes, but using it for object placement in CXL tiering raises additional questions. First, the managed runtime only manages virtual memory and is typically unaware of the physical backing of each virtual page, i.e., whether it resides in local memory or CXL memory. As a result, simply moving objects within the heap does not necessarily place hot objects in the fast tier. Therefore, the runtime must either know or coordinate with the system that controls the physical placement of heap pages. Second, although most objects are smaller than a page, large objects that span multiple pages still exist, with arrays being a common example. Relocating such objects as a whole is, in fact, coarser-grained than page-level management, so large objects require different handling.

\para{\emph{Our solution}:} \name uses a \textit{hybrid relocation approach} that combines runtime-level object placement with page-based CXL management to address these challenges. This approach separates responsibilities between the managed runtime and the page-based system: the managed runtime compacts hot objects into contiguous virtual pages, which we call \textit{hot-object compaction}, while the page-based system migrates the corresponding physical pages between tiers. This design successfully places hot objects in fast-tier memory without requiring the runtime to know the physical backing of heap pages: once hot objects are compacted into a small set of virtual pages, any reasonable page-based system will identify the corresponding physical pages as hot and thus migrate them to the fast tier. This arrangement also naturally handles large objects at page granularity; \name leaves such objects in place on the heap and lets the underlying page-based system manage their physical placement across tiers.

To perform hot-object compaction, \name reuses the GC's relocation machinery in two ways. First, during normal GC relocation, where the GC moves live objects to compact the heap, \name additionally checks each object's hotness counter and redirects hot objects to a designated hot-object space. Cold objects continue to be consolidated in the original heap-compaction destination, thereby separating hot and cold objects. Second, when normal GC cycles are too infrequent, \name can invoke a separate relocation phase dedicated to hot-object compaction, reusing the same object-moving mechanism without performing normal heap compaction. We describe the precise policies governing these relocation decisions in the next section.

\subsection{Object Relocation Policies}
\para{\emph{Challenge}:} During hot-object compaction, the runtime must carefully decide which objects to relocate. Selecting too many objects can dilute the hotness density of the resulting pages, while selecting too few objects can leave too few hot pages for page-based migration to populate fast-tier memory. 

Furthermore, hot-object compaction can unnecessarily increase GC relocation time under today's \textit{region-based} GC. Relocating all live objects in the heap during every GC cycle would be too expensive, so modern garbage collectors~\cite{yang2022deep,oracleHotSpotVirtual} typically divide the heap into fixed-size regions and relocate objects in only selected regions, such as those with high fragmentation. This region-based design complicates hot-object compaction: relocating hot objects from a region unnecessarily requires relocating other cold live objects in the same region when the region would not otherwise be selected by the normal GC policy. As a result, the runtime must ensure that the improvement made by relocating hot objects in a region outweighs the extra relocation overhead.

\para{\emph{Our solution}:} 
\name introduces two policies to control object-placement quality and GC relocation time (\S\ref{subsec:design-coloc}).

To improve hotness density, \name dynamically determines a \textit{hotness cutoff} and relocates only objects whose hotness counters exceed the cutoff. During the GC object-graph scan phase, which precedes relocation, \name reads the hotness counters and builds a global view of object hotness. It then determines a cutoff such that only sufficiently hot objects are relocated to fill the available fast-tier capacity. 

To control relocation overhead in region-based GC, \name selects regions for hot-object compaction based on the ratio of hot objects within each region. To evaluate the benefit of relocating hot objects from a region, \name calculates the total bytes of hot objects, i.e., objects whose counters exceed the cutoff, in that region at the start of the GC relocation phase and compares it against adjustable watermarks. This policy selects regions for additional relocation only when the expected benefit is sufficiently high, thereby bounding the GC overhead introduced by hot-object compaction.

\subsection{System Overview}
\begin{figure}[t]
    \centering
    \includegraphics[width=0.95\columnwidth]{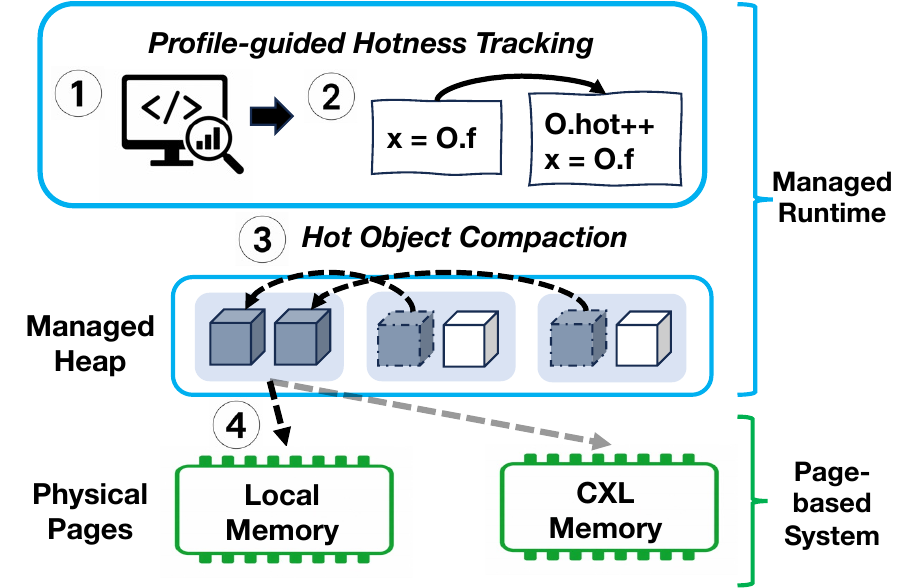}
    \vspace{-0.1in}
    \caption{\name system overview. Cubes represent objects; shaded cubes indicate hot objects.}
    \vspace{-0.2in}
    \label{fig:overview}
\end{figure}
\name operates in the following logical steps, as shown in Figure~\ref{fig:overview}. \circled{1}: For profile-guided hotness tracking, \name's online profiler (\S\ref{subsec:design-profiler}) collects L3 cache-miss samples using PEBS at a low sampling rate and identifies delinquent instructions to instrument. \circled{2}: The profiler passes the identified instructions to the JIT compiler, which in turn instruments those instructions with hotness-tracking logic that increments an object hotness counter (\texttt{O.hot++} in the figure). As the program runs, the counters in object headers are updated. \circled{3}: Hot-object compaction occurs as part of GC or is invoked separately when \name determines additional compaction is needed based on the hot object distribution (\S\ref{subsec:design-coloc}). It begins by scanning the object graph and reading the hotness counters to dynamically determine the hotness cutoff based on the hotness histogram and the capacity of local memory. During normal GC, it then selects regions in which objects are relocated, in addition to GC's original relocation targets. Finally, \name scans target regions and relocates objects whose counters exceed the cutoff into dedicated regions (\S\ref{subsec:design-coloc}). \circled{4}: The underlying page-based system then detects that pages in the target regions have become hot and migrates them to local memory by updating the page table (arrows in the figure).

\section{Design}
\label{sec:design}
We next elaborate on the main components of \name's managed runtime: profile-guided hotness tracking in \S\ref{subsec:design-profiler} for online profiling and \S\ref{subsec:design-hotness-tracking} for hotness-tracking logic, and policies for hot-object compaction in \S\ref{subsec:design-coloc}. We use the JVM when needed to give concrete examples throughout this section because our prototype is implemented in the JVM (\S\ref{sec:impl}), but the design principles are not JVM-specific.

\subsection{Online Profiling}
\label{subsec:design-profiler}
\noindent The goal of the online profiler is to identify the instructions that are the source of the majority of L3 cache misses during runtime. For this purpose, the profiler should accurately identify all delinquent instructions with low overhead. %

To meet these requirements, \names profiler uses hardware-assisted sampling, PEBS in our prototype, to collect L3 cache miss samples, which contain the instruction pointer (IP) of the triggering instructions. Unlike L1/L2 misses, L3 cache misses often stem from a small number of instructions. As a result, the profiler can identify these delinquent instructions with high accuracy and low overhead using a relatively low sampling rate.
To integrate profiling into the managed runtime, \name sets up PEBS events during the runtime initialization, which involves creating a ring buffer for each CPU core to record L3 miss samples. Then, it spawns a buffer-monitoring thread that regularly wakes up and consumes the PEBS samples from the buffers. The thread reads the IP field of the samples and records the number of appearances of each instruction in a hashmap.

To maintain a stable delinquent-instruction list while still capturing recency, \names profiler determines delinquent instructions based on the exponential moving average of sample counts over windows instead of simply accumulating the counts. 
Finally, the profiler computes the ratio of each instruction's exponential moving average of the sample counts to the sum across all instructions, selecting those with counts higher than a threshold as delinquent instructions. The decaying period is based on the number of L3 cache miss samples (see \S\ref{sec:impl} for the configuration in our prototype).

\subsection{Hotness Tracking}
\label{subsec:design-hotness-tracking}

\subsubsection{Instrumenting Delinquent Instructions}
In JIT-based managed runtimes, programs are first compiled to a language-level IR and later compiled into optimized native machine code by the JIT compiler. To insert hotness-tracking logic, \name first maps each identified delinquent instruction back to the corresponding IR instruction and requests recompilation of the affected code. In the JVM, this intermediate form is Java bytecode. \name maps the IPs of delinquent instructions back to Java bytecodes and marks those bytecodes as instrumentation targets. It then uses deoptimization, which invalidates the existing optimized machine code and triggers recompilation of the target bytecodes, to insert hotness-tracking logic.

This approach requires careful handling of the function’s call context. Whether an instruction is delinquent often depends on its position in the call stack, so indiscriminately marking IRs in frequently used methods as instrumentation targets can introduce unnecessary overhead. This issue is exacerbated by the frequent use of many short utility functions in diverse contexts. To address this, when \name observes a delinquent instruction, it records the corresponding caller context along with the marker. In the JVM, this context is represented as pairs of bytecode indices and callers in the call stack. When the JIT compiler encounters a marked instruction during parsing, it checks the current callers in the call stack against the recorded pairs and performs instrumentation only if all matches are confirmed.

\begin{figure}[t]
    \centering
    \includegraphics[width=0.9\columnwidth]{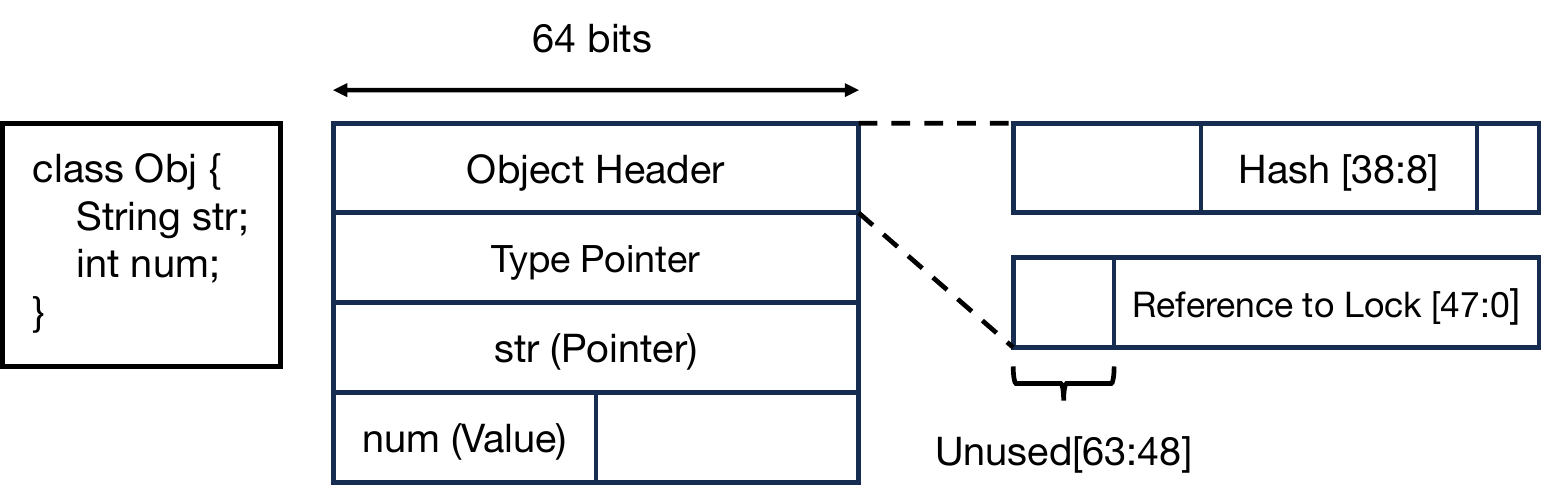}
    \caption{JVM object layout in a 64-bit system. The upper 16 bits of the header are unused.}
    \vspace{-0.15in}
    \label{fig:object}
\end{figure}

\begin{lstlisting}[
style=gasstyle,
float,
caption={\names hotness tracking logic in x86 assembly. It reads the counter field in the header, increments it, and writes it back. If the counter reaches the limit, the header update is skipped. \texttt{movzwq} and \texttt{movw} represent 16-bit \texttt{mov} operations.},
label={listing:logic}
]
inc_counter(Register scr, Address header_addr) {
    movzwq %
    cmp %
    je equal // if (scr == 2^16-1), skip
    inc %
    movw 0x6(obj),%
equal:
    ... // delinquent load instruction
}
\end{lstlisting}

\begin{table}[t]
\footnotesize
\centering
{%
\begin{tabular}{|l|l|l|}
\hline
\textbf{Bytecode} & \textbf{Access Type}             & \textbf{Example Code}  \\ \hline
getfield           & Fields                 & \texttt{x = O.f}                \\ \hline
checkcast          & Type pointer           & \texttt{String x = O.get(k)}    \\ \hline
instanceof             & Type pointer           & \texttt{if (x == null) \{...\}}\\ \hline
invokevirtual      & Type pointer           & \texttt{O.put(k,v)}             \\ \hline
arraylength        & A fixed field & \texttt{if (arr.length == 5) \{...\}}   \\ \hline
\end{tabular}
}
\caption{The list of Java bytecodes profiled by \name.}
\vspace{-0.15in}
\label{table:bc}
\end{table}

\para{Hotness Tracking Logic} The efficiency of hotness tracking logic is vital in \names design. 
\name keeps this logic lightweight by reusing object-header bits and minimizing the number of inserted instructions. To record the accesses by delinquent loads, \name uses unused bits in the object header (top 16 bits in the case of the JVM, as shown in Figure~\ref{fig:object}). This approach avoids the space overhead of separate counters and prevents additional cache misses that could arise if counters were stored in separate memory.

Listing~\ref{listing:logic} shows \names hotness tracking logic. We highlight two aspects here. First, this implementation is highly efficient because it spans only 5 lines of assembly code, and since the branch direction changes only when the counter exactly reaches its limit, the control flow overhead is completely mitigated by the branch predictor. Second, this logic assumes that a scratch register and the address of the object header are provided. 
We take the JVM's JIT compiler as an example to show how this requirement can be met during code generation. For the scratch register, the register used to store the result of the load can be freely used at the time of hotness tracking. For the object header, we observe that the memory operand of load instructions is offset by a fixed number of bytes from the object header, depending on the bytecode (Table~\ref{table:bc}). For example, in the \texttt{getfield} bytecode, the memory operand of the load instruction takes the form \texttt{[REG + offset]}, where \texttt{REG} contains the address of the object header, and \texttt{offset} specifies the displacement of the target field relative to the header. %
\name similarly modifies the compilation of each bytecode, ensuring that the header address is correctly provided to the hotness tracking logic.

\begin{lstlisting}[
  style=gasstyle,
  float={!tp},
  caption={Sampling increments a per-thread counter at every invocation (Top), while periodic activation uses the same counter value until a background thread periodically increments it (Bottom). \texttt{T} represents thread-local storage. },
  label={listing:sampling-vs-pa}
]
// Sampling: application thread
if (T.sampling_counter++ == N) { inc_counter(); }
## a delinquent load instruction ##

// Periodic activation: application thread
if (T.activation_counter == N) { inc_counter(); }
## a delinquent load instruction ##

// Periodic activation: background thread
while (true) { T.activation_counter++; sleep(1ms); }
\end{lstlisting}

\subsubsection{Refining Hotness Tracking}
We next present several complementary mechanisms that improve the tracking efficiency and accuracy. First, although lightweight, our hotness tracking mechanism can still incur noticeable overhead when fully turned on. \name thus provides a knob to balance the overhead-accuracy trade-off by controlling how often the tracking logic is enabled. Second, \name periodically refreshes the object hotness counters to reflect the recency of accesses and handle saturated counters.
Lastly, \name efficiently resolves the conflicts on the object header between the hotness tracking mechanism and locking primitives. 

\para{Periodic Activation} One conventional way to provide a knob for the overhead-accuracy trade-off is to employ uniform sampling, tracking only one event out of $N$, where $N$ is the sampling period, by adding a conditional check and a sampling counter. However, the check for sampling is entirely unpredictable by the branch predictor, adding significant misprediction overhead. Also, sampling loses the ability to track objects with locality. For example, objects associated with a single key-value pair in a hashmap are typically allocated contiguously and used at the same time. Tracking their usage through uniform sampling can produce inconsistent hotness values for these related objects, leading to relocation decisions that disrupt their locality.

To address these issues, \name employs \textit{periodic activation} of the hotness tracking logic as the overhead-accuracy knob. Listing~\ref{listing:sampling-vs-pa} highlights the difference between sampling and periodic activation. Periodic activation also adds a conditional check before invoking the hotness tracking logic, similar to sampling. However, the condition is updated periodically by a background thread, rather than by the application thread, and much less frequently (e.g., every 1 ms). This approach enables the branch predictor to handle the condition check correctly in nearly all cases and allows the hotness tracking logic to continue tracking objects with locality.

\para{Refreshing Counters} 
Our hotness tracking logic is intentionally designed to be simple: it only increments the counter monotonically for efficiency. As a result, it fails to capture the recency of accesses and react to hotness shifts. To address this issue, when the online profiler reaches its L3-miss sample threshold (\S\ref{subsec:design-profiler}), \name initiates an object-graph scan and decays the hotness counters in objects.

\para{Handling Locking Primitives} 
This issue is specific to managed runtimes that use the object header as locking metadata. In the JVM, object-level locking relies on internal locking primitives that store locking metadata in the object header. When a thread locks an object, it installs this metadata using a compare-and-swap (CAS) operation to handle contention. The CAS operation is expected to fail only when another thread is concurrently modifying the lock state. However, because \name's hotness-tracking logic also modifies the object header, it can cause false CAS failures.

Modifying the hotness tracking logic to acquire a lock or use expensive atomic operations to resolve this issue would significantly harm its efficiency. To address this, \name adopts an optimistic retry approach, assuming that false failures are rare. It keeps the hotness tracking logic intact and instead adds a check-and-retry after each CAS operation for locking. When a CAS operation fails, the thread checks whether the lower 48 bits of the object header remain unchanged before and after the CAS attempt. If no changes are detected, the failure is classified as false, and the thread retries the CAS operation. With this optimization, we confirmed that adding hotness tracking logic to concurrent data structures did not show any noticeable overhead.

\subsection{Policies for Hot-Object Compaction}
\label{subsec:design-coloc}
We next elaborate on the two policies governing the hot-object compaction process. %

\para{Determining Hotness Cutoff}
\name sets the hotness cutoff to approximate an ideal capacity-aware placement policy: place objects in the fast tier in descending order of hotness until the fast-tier capacity is filled. To approximate this policy efficiently, \name computes a histogram of object hotness during the GC object-graph scan and derives the cutoff from this histogram and the available local memory size. As the GC traverses live objects, \name reads each object's hotness counter from its header. \name then maps the counter value to a histogram bin and adds the object's size to that bin. After the scan completes, \name computes a cumulative sum over the bins in descending order of hotness. If the cumulative sum first exceeds the local memory size at bin $i$, then objects in bins $i+1$ and higher are classified as hot for the current GC cycle. \name uses exponential bins, where the $i$-th bin contains objects with counter values in $[2^i, 2^{i+1})$, allowing it to make fine-grained decisions with a small number of bins.

\para{Region Selection}
The region selection policy in the existing region-based GC is designed to reduce heap fragmentation. Therefore, hot-object compaction requires \name to select additional regions using a different criterion: whether compacting hot objects from the region is likely to improve fast-tier hit ratio enough to justify the extra relocation work.

The key observation is that selecting regions with either too many or too few hot objects is undesirable. Regions that are already densely populated with hot objects offer little additional benefit from compaction. Conversely, regions with too few hot objects are not worth processing because region scanning and relocation setup impose fixed costs; selecting such regions would add relocation overhead without substantially improving hot-object density.

Based on this observation, \name adopts a region-selection policy based on two configurable watermarks: low and high. \name selects only regions whose hot-object ratio, measured in bytes, falls between these two watermarks.
This policy balances fast-tier utilization against relocation overhead by avoiding low-benefit regions at both extremes. It is also efficient to implement by recording the number of hot bytes in each region during the object-graph scan. In our prototype, we use fixed values for watermarks (\S\ref{sec:impl}) because they are effective across applications in our evaluation. We leave the dynamic adjustment of these watermarks to future work.

\para{Additional Relocation Phases}
While hot-object compaction piggybacks on GC relocation, normal GC cycles may not occur frequently enough to rebalance hot objects as their hotness changes. To address this, in addition to normal GC runs, \name computes each region's hot-object ratio after the object-graph scan used to refresh hotness counters and invokes an additional relocation phase dedicated to hot-object compaction if enough regions fall between the watermarks.

\subsection{Discussion}
\label{subsec:discus}
\para{Applicability to Other Managed Languages} While implementation details differ across runtimes, \name{}'s core design principles are not JVM-specific. \name{}'s design relies on three runtime capabilities: object metadata for hotness tracking, JIT compilation for dynamic instrumentation, and moving garbage collection for hot-object compaction. These capabilities are available not only in HotSpot JVM and JVM-based languages such as Java, Kotlin, and Scala, but also in other managed runtimes such as .NET CLR~\cite{dotnet-object-header,dotnet-managed-execution,dotnet-gc}, PyPy~\cite{pypy-gc,pypy-jit}, and modern JavaScript engines~\cite{v8-gc,v8-maglev,spidermonkey-gc}.

\para{Target Applications} \name targets memory-intensive applications whose hot sets evolve over minutes or hours, as is common in request-serving workloads~\cite{yang2021large,atikoglu2012workload,tang2017popularity} and assumed in much of the tiered-memory literature~\cite{lee2023memtis,maruf2023tpp,vuppalapati2024tiered}. Our evaluation (\S\ref{subsec:factor}) shows that, by leveraging the JVM’s existing GC infrastructure, object-graph scanning and hot-object relocation can be performed at the scale of 30 million objects in tens of seconds, which is sufficiently fast for \name to adapt to hotness shifts and relocate objects at these timescales. Extreme workloads in which the hotness of millions or billions of objects changes every few seconds are outside our scope.

\para{Off-heap Memory}
Off-heap memory, such as native allocations and buffers for I/O, does not participate in GC-based relocation. Such memory can still be managed at least at page granularity by the underlying page-based system.

\section{Implementation}
\label{sec:impl}
We prototyped \name by extending OpenJDK 21~\cite{openjdk}, the most widely used open-source managed runtime. 

We implement the online profiler as a native background thread within the JVM. We use a PEBS sampling rate of 1/2000, apply a decay ratio of 1/2 to sample-appearance counts and hotness counters every 1 million L3-miss samples, and use 1\% as the cutoff for identifying delinquent load instructions. While this choice of fixed parameters works well in our evaluation, adaptive aging or different recency-frequency ratios may better handle more complex access patterns~\cite{megiddo2003arc,lee1999existence}. We leave the exploration of such policies to future work.

To implement hotness tracking, we modified the C2 JIT compiler~\cite{paleczny2001java}, which generates the most optimized code, in two ways. First, we repurposed the \texttt{ScopeDesc} data structure, which maintains bytecode and inlining context for debugging and deoptimization, to map the IP of each delinquent load back to its original bytecode. Second, we altered the bytecode parsing and code generation phases to insert our instrumentation logic. Specifically, we modified the C2 compiler as follows: when the compiler parses a bytecode mapped to a delinquent load, it generates a special \texttt{Load} IR node. This node is treated identically to a regular \texttt{Load} node throughout compilation, and the compiler additionally injects our hotness tracking logic before the load instruction when generating machine code for this node during final code generation. This approach enabled instrumentation without interfering with existing C2 compiler optimizations.

We implemented hot-object compaction by extending ZGC~\cite{yang2022deep,openjdk439Generational}, a modern pauseless, region-based garbage collector. ZGC partitions the heap into 2 MB regions, and its full GC already includes the three phases \name relies on: object-graph traversal, region selection, and relocation. We modified these three phases as described in \S\ref{sec:design}, which naturally enables hot-object compaction during normal full-GC passes. \name can also perform hot-object compaction when invoked by the object-graph scan: in this case, it reuses the full-GC pass structure but skips the normal heap-compaction relocation phase, performing only the relocation work needed for hot-object compaction. We use fixed low and high hot-object-ratio watermarks of 5\%/50\%.

\section{Evaluation}
\label{sec:eval}
In this section, we first present our evaluation setup (\S\ref{subsec:eval-setup}) and then answer three key questions: (i) How much does \name reduce application slowdown compared to page-based systems? (\S\ref{subsec:eval-performance}); (ii) Does \name improve fast-tier hit ratio? (\S\ref{subsec:eval-hit-ratio}); and (iii) Do \names components achieve their goals with low additional runtime overhead? (\S\ref{subsec:factor}).

\subsection{Evaluation Setup}
\label{subsec:eval-setup}

\para{Methodology and Testbed} We use a dual-socket system to emulate a CXL-based tiered memory architecture, following the methodology in prior work~\cite{lee2023memtis,amaro2023logical,song2025hybridtier}. Application and
runtime threads are pinned to the local socket using \texttt{taskset}. We test three different tiering ratios, where 10\%, 20\%, or 50\% of memory is local, and the tiering ratios are enforced using the local-memory ballooning
mechanism used in prior work~\cite{lee2023memtis}. We conduct our experiments on a
CloudLab c6420 machine~\cite{duplyakin2019design} equipped with Intel Xeon Gold
6142 CPUs and 192~GB DDR4 memory per socket.

\begin{figure}[t]
    \centering
\includegraphics[width=\columnwidth]{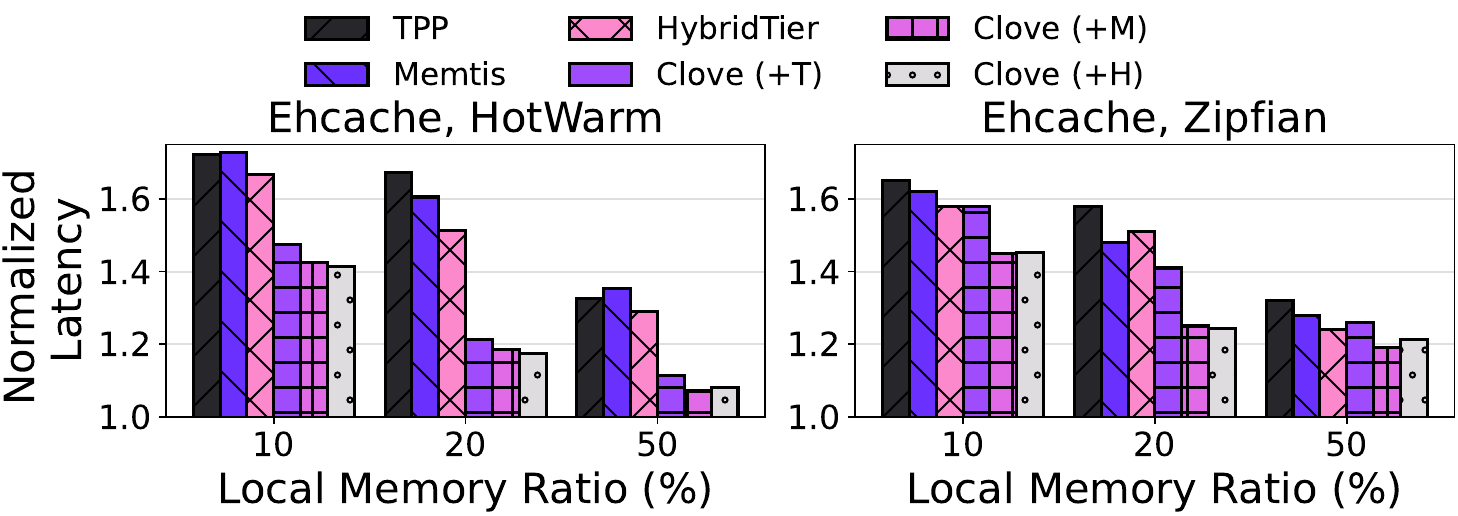}
        \vspace{-0.25in}
        \caption{Synthetic workload performance. Latency is normalized to the all-local case (lower is better). "\name (X)" represents \name using X as the underlying page-based system.}
        \vspace{-0.3in}
    \label{fig:eval-synt-perf}
\end{figure}

\begin{figure*}[t]
    \centering
        \includegraphics[width=0.95\textwidth]{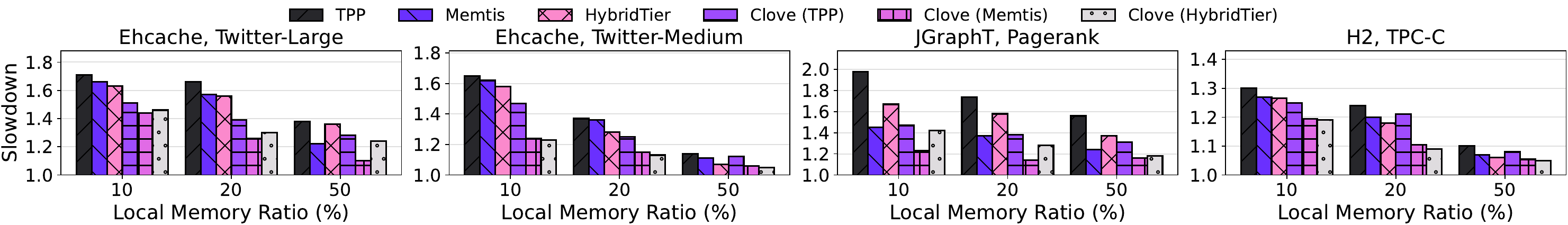}
        \vspace{-0.1in}
    \caption{Performance on real-world workloads. Slowdown is measured relative to the all-local case (lower is better). "\name (X)" represents \name using X as the underlying page-based system.}
    \label{fig:eval-real-perf}
\end{figure*}

\para{Systems}
We evaluate three open-source, page-based tiered-memory management systems---TPP~\cite{maruf2023tpp}, Memtis~\cite{lee2023memtis,githubMemtis}, and HybridTier~\cite{song2025hybridtier,githubHybridTier}---as both baselines for comparison and as the underlying page migration systems for \name. TPP uses NUMA hint faults and page list scans for promotion/demotion, representing the state of the art among page-fault--based systems. Memtis and HybridTier use PEBS sampling to distinguish hot from warm pages, making them adaptable to workloads with diverse access hotness. The two systems focus on different aspects: Memtis introduces dynamic hugepage splitting to reduce space waste, while HybridTier employs counting Bloom filters to lower metadata space overhead and cache misses in hotness tracking.

We evaluate \name integrated with each of the three baseline page-based systems.
Since \name is designed to interoperate with any page-based system, we integrated \name with baseline systems without modifying their code.
We enable Transparent HugePages~\cite{redhat52xA0HugePages} in all experiments and Memtis's hugepage split, as they consistently improve performance across workloads. We set the JVM heap size to be 20\% above the memory footprint of each workload to avoid out-of-memory. Other system parameters, including those mentioned in \S\ref{sec:impl}, remain at their defaults unless specified.

\para{Applications} We evaluate \name and the baseline systems with three memory-intensive Java applications, similar to the native applications used in prior studies~\cite{lee2023memtis,vuppalapati2024tiered,maruf2023tpp,song2025hybridtier}. \textit{Ehcache}~\cite{ehcacheEhcache} is a widely used key-value cache library. We use a YCSB-C--style benchmark~\cite{cooper2010benchmarking} with steady QPS on Ehcache and measure query latency. \textit{JGraphT}~\cite{jgraphtJGraphT,michail2020jgrapht} is a pure-Java graph algorithm library, where we evaluate the PageRank implementation and measure completion time. Lastly, \textit{H2}~\cite{h2databaseDatabaseEngine} is a relational database; we run it in in-memory mode to focus on scenarios where memory speed is the performance bottleneck. We also measure query latency for H2.

\subsection{End-to-End Performance}
\label{subsec:eval-performance}
This subsection answers Q1: whether \name reduces end-to-end slowdown compared to page-based systems.

\para{Synthetic Workloads} \label{subsec:eval-synt}
We first evaluate \name using Ehcache with synthetic workloads, as they provide interpretable access patterns with object-level skewness. Each workload uses 4 KB values (much smaller than the 2 MB page) with a total footprint of 100 GB.
We consider two skewed access distributions: HotWarm and Zipfian (0.99). In HotWarm, 20\% of keys account for 90\% of accesses, and accesses within the hot set are uniform. Thus, object-level management should be able to capture 90\% of accesses in the fast tier once the fast tier reaches 20\% of the total footprint. In Zipfian, accesses are concentrated on a smaller set of very hot objects, so even page-granularity management can capture many hot objects despite space waste. Consequently, the relative benefit of object-level management is expected to be smaller and to reach diminishing returns more quickly as local-memory capacity increases. Ehcache maintains a large reference array for the hashmap and stores payload and metadata objects for each key--value pair. Because the hashmap array is relatively small and contiguous, overall performance is dominated by the placement of the objects associated with hot keys.

Figure~\ref{fig:eval-synt-perf} shows that \name consistently outperforms page-based baselines. In the HotWarm distribution, even though hot objects fit within 20\% of total memory, page-based systems cannot place them in a contiguous space, leading to significant slowdowns even with 50\% local memory. Memtis and HybridTier perform slightly better than TPP by distinguishing hot from warm pages, but still fall short of ideal. In contrast, \names hot-object compaction ensures hot objects are packed contiguously, so when local memory starts exceeding the hot-object footprint (20\%), most cache misses are served locally. This yields a 29--59\% latency reduction at a 20\% local-memory ratio compared to baselines. Under the Zipfian distribution, the compaction benefits are slightly reduced but remain evident at the 10\% and 20\% local memory ratios. The performance gap narrows as more local memory is provided, since page-based systems eventually capture hot objects despite space waste.

\para{Realistic Workloads} We next evaluate \names effectiveness on real-world workloads that exhibit skewed access patterns. Since our testbed has 192 GB of local memory, we scale each workload so that its footprint is of comparable size. For Ehcache, we use two Twitter production traces~\cite{githubTwittercachetrace} with footprints of 160 GB and 120 GB, which we denote Twitter-Large and Twitter-Medium to highlight their average payload sizes (44~KB, 3~KB). For PageRank, we use a benchmark graph from the GAP benchmark~\cite{beamer2015gap}, which has a 120 GB footprint, following prior work~\cite{vuppalapati2024tiered,lee2023memtis}. For H2, we run TPC-C~\cite{tpcTPCCHomepage} and adjust record sizes so that the total footprint is 120 GB.

\para{Ehcache} The first two columns of Figure~\ref{fig:eval-real-perf} show the slowdown of the systems on Ehcache with the Twitter production traces. Ehcache's memory usage characteristics are described above. The performance trend resembles that of the Zipfian distribution, as the Twitter traces are also known to exhibit Zipfian distributions, with slightly higher skewness than 0.99. As a result, \name reduces slowdown relative to all-local by 29--59\% (Twitter-Large) and 43--63\% (Twitter-Medium). The gap between \name and the baselines is larger in Twitter-Medium, since its hottest working set---accounting for 80\% of accesses---fits within 10\% of local memory. This gap narrows as more local memory is provisioned.

\para{JGraphT} The third column of Figure~\ref{fig:eval-real-perf} shows the slowdown in PageRank completion time. Memory usage in JGraphT is dominated by three types of objects: Node objects storing keys and ranks, a hash table mapping Nodes to their adjacency lists, and the adjacency lists themselves. The first two types are relatively small but are the hottest, since they are accessed in every iteration. In contrast, adjacency lists account for most of the footprint, with hotness proportional to each node's degree. TPP fails to differentiate the hottest Node and hash table objects from adjacency lists, leading to the worst performance. Memtis and HybridTier capture the pages containing these hot objects more effectively, narrowing the gap. In contrast, \name also identifies the hottest adjacency lists and compacts them, yielding a 47--84\% improvement over the baselines.

\para{H2} As a B-tree--based DBMS, H2's memory footprint is primarily composed of B-tree nodes and record objects (arrays of columns). TPC-C is well-known to have skewed record accesses, with 75\% of accesses going to 20\% of records~\cite{lam2024accelerating,leutenegger1993modeling}. This naturally creates hot record objects that \name can exploit for performance improvement. As a result, \name reduces slowdown in H2 by 22--47\% across different tiering ratios, as shown in the last column of Figure~\ref{fig:eval-real-perf}. The absolute performance gap between systems is smaller in H2 than in the other applications, since H2 also spends time on CPU-intensive tasks such as transaction processing and indexing.

\subsection{Does \name Improve Fast-tier Hit Ratio?}
\label{subsec:eval-hit-ratio}
\begin{figure}[t]
    \centering
    \includegraphics[width=\columnwidth]{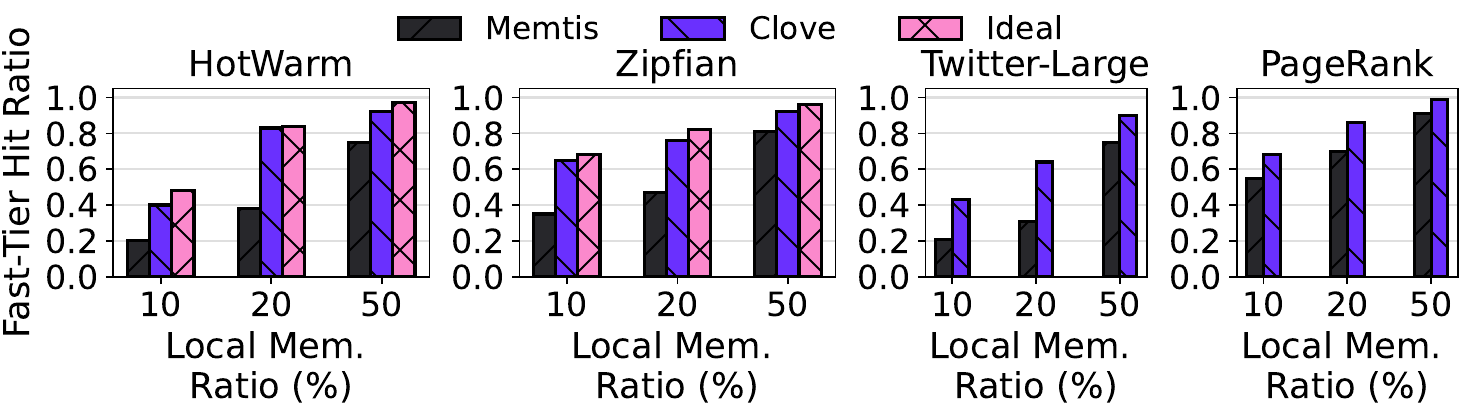}
    \vspace{-0.3in}
    \caption{Local memory hit ratio in synthetic and realistic workloads. Application names are omitted.}
    \vspace{-0.2in}
    \label{fig:eval-synt-hr}
\end{figure}

To answer Q2, we measure the local memory hit ratio using PEBS samples under two setups: Memtis and \name integrated with Memtis. For synthetic workloads, we also include Ideal, where we modify the workload to place keys in strict order of hotness so that the resulting fast-tier placement matches the object-level oracle placement described in \S\ref{subsec:intrapage}; we then measure the resulting hit ratio.
The first two columns in Figure~\ref{fig:eval-synt-hr} show that \name achieves 11--45 percentage points higher hit ratios than Memtis. Moreover, \name's hit ratio is only 4--5 percentage points lower than Ideal, showing that \name achieves a near-optimal hit ratio. The two rightmost columns show that a similar trend appears in realistic workloads.

\subsection{Factor Analysis}
\label{subsec:factor}
This subsection answers Q3 by showing that \names end-to-end benefits stem from careful design choices that achieve their goals while bounding the overhead of its main components. We use "\name (Memtis)" as a representative case.
\begin{figure}[t]
    \centering    \includegraphics[width=   \columnwidth]{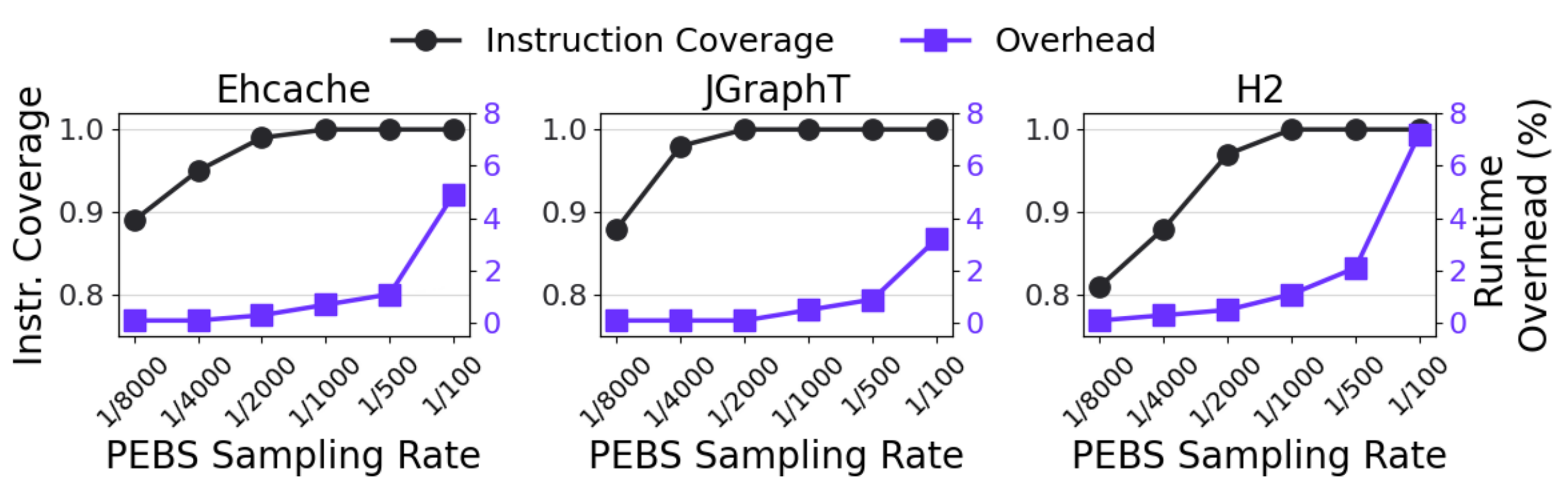}
    \vspace{-0.18in}
    \caption{Instruction coverage and runtime overhead of the online profiler with different PEBS sampling rates.}
    \vspace{-0.2in}
    \label{fig:eval-profiling-overhead}
\end{figure}

\begin{figure}[t]
    \centering
        \includegraphics[width=\columnwidth]{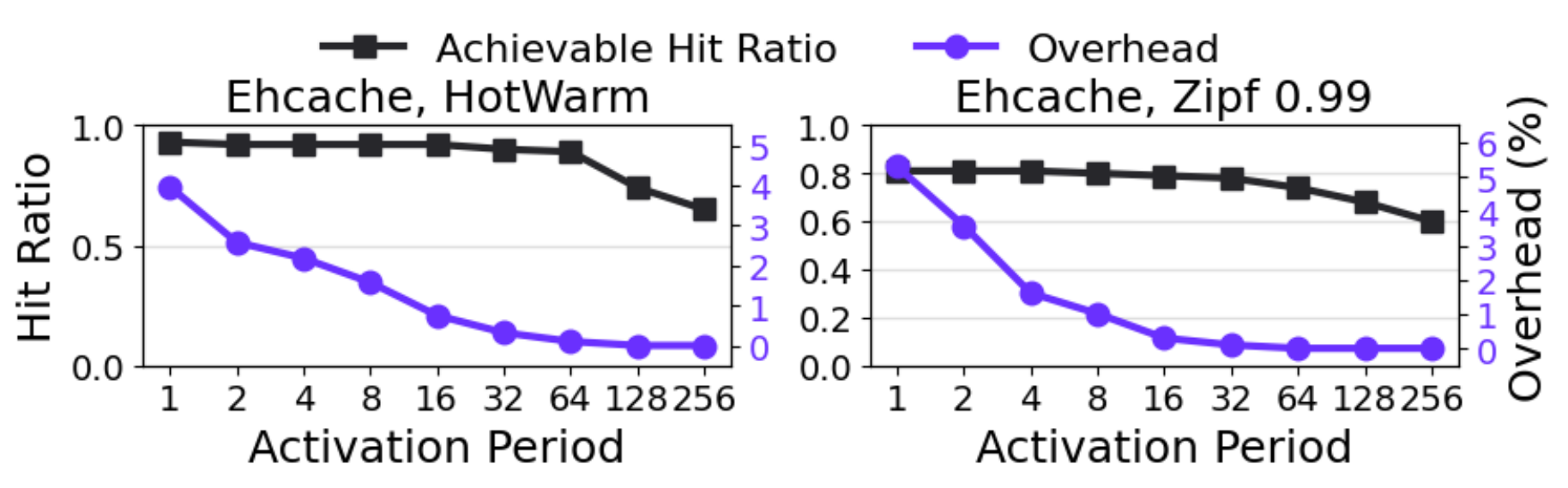}
        \vspace{-0.1in}
    \caption{The effect of periodic activation.}
    \vspace{-0.2in}
    \label{fig:eval-inst-overhead}
\end{figure}
\subsubsection{Online Profiler} We evaluate whether \names online profiler identifies delinquent loads with low overhead. We compare the delinquent-instruction list identified at each PEBS sampling rate against the list identified at a 1/100 sampling rate, and measure profiling overhead at each rate. Figure~\ref{fig:eval-profiling-overhead} shows that instruction coverage quickly saturates near 100\% around a 1/2000 sampling rate, while runtime overhead remains below 1\% across applications. This shows that PEBS can identify delinquent loads accurately with low overhead.

\subsubsection{Hotness Tracking} 
We evaluate whether \name's hotness tracking achieves high accuracy with low runtime overhead through lightweight logic and periodic activation. For accuracy, we measure the local memory hit ratio at a 20\% local memory ratio. For overhead, we run \name in the all-local configuration with only profiling and instrumentation enabled, then measure the slowdown. We use synthetic workloads to isolate instrumentation overhead from dynamic workload behavior.

Figure~\ref{fig:eval-inst-overhead} shows the results. The activation period $N$ denotes that hotness tracking is enabled for 1 ms every $N$ ms. We highlight three points. First, even without periodic activation, overhead remains modest---about 5.5\% and 4\% in each case---because the hotness-tracking logic is lightweight. Second, periodic activation effectively eliminates this overhead with virtually no loss in accuracy for $N = 8, 16, 32$, demonstrating the value of this technique. Finally, accuracy begins to degrade only when the activation period becomes very large ($N \geq 64$), suggesting that \name can accommodate scenarios with higher instrumentation overhead.

\subsubsection{Hot-Object Compaction}
Lastly, we evaluate the effectiveness of and overheads related to hot-object compaction. We separate the analysis into three parts: GC overhead, whether periodic activities help \name adapt to dynamic workloads, and the effectiveness of the compaction policies.

\begin{figure}[t]
    \centering    \includegraphics[width=\columnwidth]{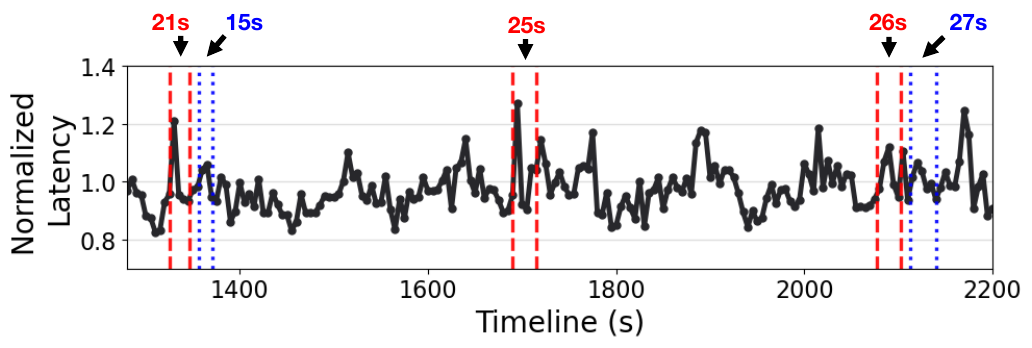}
        \vspace{-0.2in}
    \caption{Normalized request latency during GC activities introduced by hot-object compaction. Red dashed intervals: object-graph scanning; blue dotted intervals: additional GC relocation for the compaction (labels indicate duration). Workload: Ehcache, Twitter-Medium. Latency is normalized to the average latency over the run.}
    \vspace{-0.1in}
    \label{fig:gc-overhead-timeline}
\end{figure}
\para{\emph{GC Overhead}}
\name adds GC-related overhead from three sources: checking hotness counters during normal object-graph scans, performing extra relocation work during normal GC relocation phases, and periodically running additional object-graph scans that may trigger a dedicated hot-object compaction phase. The first overhead is negligible across applications because its CPU cost is hidden by the latency to fetch the next objects.

To evaluate the latter two overheads, we measure the frequency and duration of \name's periodic object-graph scans and dedicated relocation phases. We use Ehcache with the Twitter-Medium trace during steady state, which has the most objects among our workloads (30 million) and therefore represents the worst case for this overhead. We only report the dedicated relocation phases, but the extra relocation work added to normal GC relocation phases is similar. Figure~\ref{fig:gc-overhead-timeline} shows that the additional object-graph scans occur roughly once every six minutes and take 20--26 seconds, accounting for only a small fraction of execution time. Note that these run concurrently with application execution under ZGC, so their cost appears as cache/memory contention rather than stop-the-world pause time. During a scan, normalized latency briefly increases to 1.2$\times$ the average due to this contention, but it quickly recovers once the scan completes. The relocation phase is triggered once every two object-graph scans and takes 15--27 seconds. This relocation time remains small because \name's region-selection policy carefully limits the regions to relocate, as discussed below.

\begin{figure}[t]
    \centering
    \includegraphics[width=0.95\columnwidth]{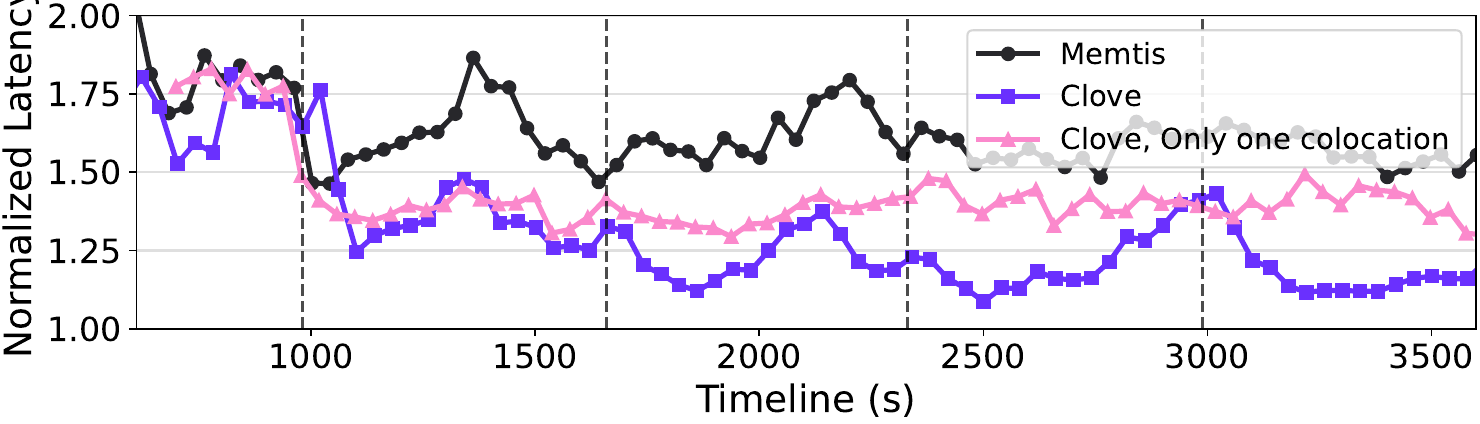}
    \vspace{-0.1in}
    \caption{\names ability to adapt to dynamic workloads. Request latency normalized to the average of the all-local case (lower is better). Local memory ratio is 20\%. Dotted lines indicate the start of the relocation phases.}
    \label{fig:eval-dynamic}
\end{figure}
\para{\emph{Adaptiveness}} Next, we demonstrate that \names periodic object-graph scans and the dedicated relocation phase help it adapt to hotness shifts. To show this point, we compare the performance of \name against Memtis and a variant of \name where hot-object compaction occurs only once, during GC at the beginning of execution.
Figure~\ref{fig:eval-dynamic} shows the performance of the three systems over time in Ehcache with Twitter-Large. Both \name and the one-time compaction variant initially outperform Memtis following the first compaction. However, the one-time variant fails to compact hot objects that appear later, resulting in worse performance than \name after the second compaction. In contrast, \name continues compacting hot objects and sustains lower latency.

\begin{figure}[t]
    \centering    \includegraphics[width=\columnwidth]{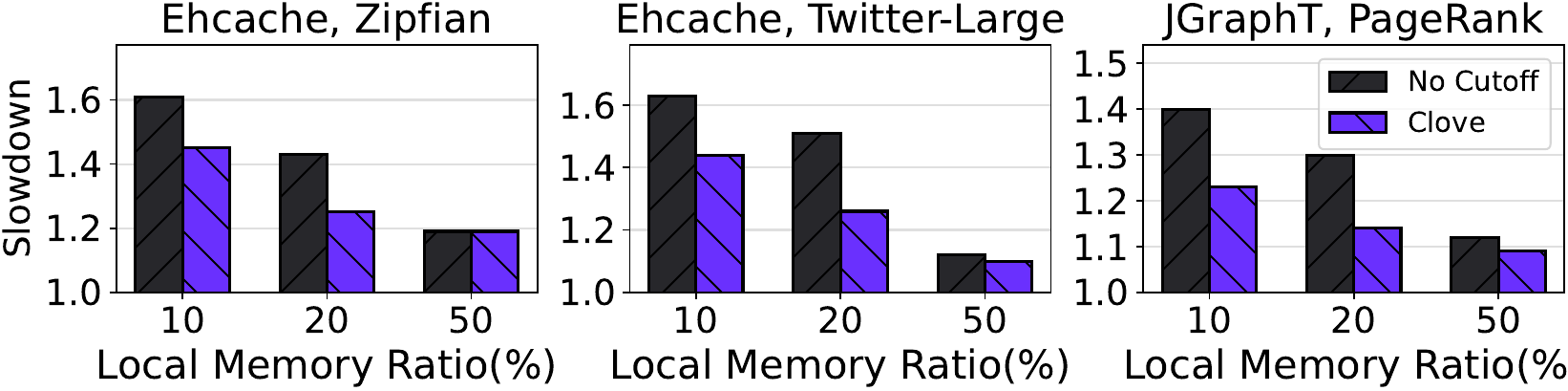}
    \vspace{-0.2in}
    \caption{The effects of the hot-object selection policy. Cutting off warm objects is necessary to achieve high local memory utilization and, in turn, reduce slowdown.}
    \label{fig:eval-cutoff}
\end{figure}
\para{\emph{Compaction Policies}} 
To assess the impact of the hotness cutoff policy on the fast-tier hit ratio, we evaluate \textit{No Cutoff}, a variant of \name that compacts all objects with non-zero counters. Results are shown in Figure~\ref{fig:eval-cutoff}. For Ehcache, No Cutoff compacts objects associated with warm keys together with hot ones, leading to degraded performance at low local memory ratios (10\%, 20\%). Similarly, in JGraphT, mixing warm adjacency lists with hot ones reduces performance to a level comparable to Memtis. These results demonstrate that the hot-object selection policy is essential for \name to achieve its performance advantage. The policy has little effect when local memory is large enough to accommodate all objects tracked in the histogram.

\begin{figure}[t]
    \centering    \includegraphics[width=\columnwidth]{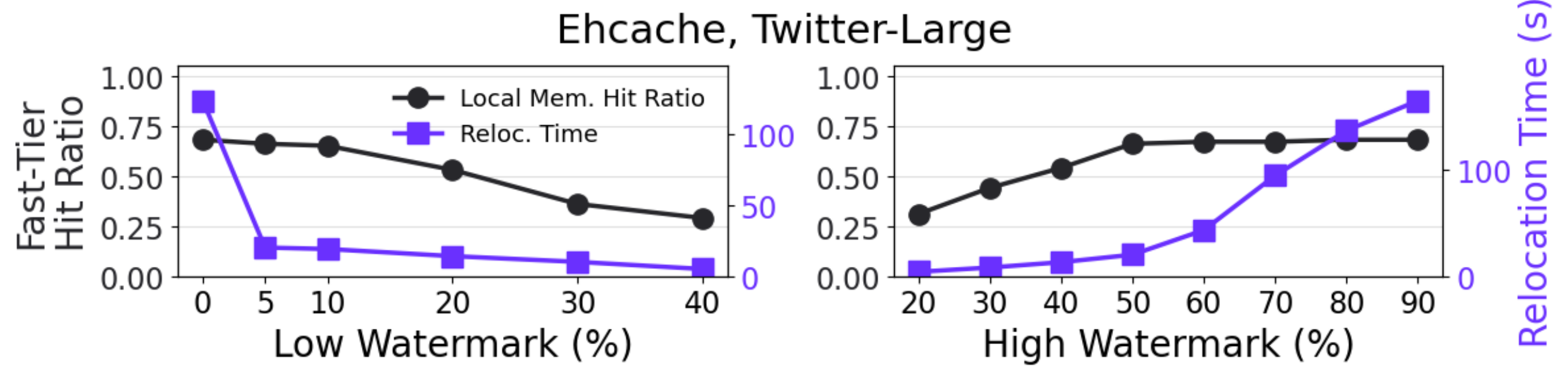}
        \vspace{-0.2in}
    \caption{The effect of region-selection watermarks. The high watermark is fixed to 50\% in the left figure; the low watermark is fixed to 5\% in the right figure.}
    \vspace{-0.2in}
    \label{fig:eval-watermark}
\end{figure}

Next, we evaluate the region selection policy. Figure~\ref{fig:eval-watermark} reports the local-memory hit ratio and additional relocation time under different low- and high-watermark settings for the same workload. When the low watermark is 0\%, any region with at least one hot object is selected, significantly increasing relocation time. Setting even a small low watermark (e.g., 5\%) reduces this overhead while preserving the local memory hit ratio. However, setting the low watermark too high prevents some hot objects from being compacted, reducing the fast-tier hit ratio. Similarly, setting the high watermark too high causes regions already dense with hot objects to be processed unnecessarily, increasing relocation time without improving the hit ratio. Setting the high watermark to 50\% reduces relocation time to a reasonable level while keeping performance stable. Overall, these results show that \names region-selection policy is important for bounding relocation overhead while maintaining performance gains.

\section{Related Work}
\label{sec:rel}
The predominant approach to CXL memory management is page-based, which suffers from intrapage hotness skew~\cite{lee2023memtis,vuppalapati2024tiered,xiang2024nomad,duraisamy2023towards,maruf2023tpp,li2023pond,zhong20242lm,autonuma,song2025hybridtier,ren2024mtm,xu2024flexmem}. 
Object-level management for tiered memory has been explored primarily in unmanaged-language systems~\cite{ruan2020aifm,wang2022memliner,wang2020semeru,amaro2020can,gu2017efficient,calciu2021rethinking,guo2023mira,tauro2024trackfm,al2020effectively,zhou2022carbink,banakar2026obase}. OBASE~\cite{banakar2026obase}, a concurrent work, applies a similar approach to the CXL context. OBASE also uses object-level reorganization as a frontend to page-based migration, similar to \name. However, OBASE targets unmanaged languages and requires developers to rewrite pointer-based data structures with special compiler annotations. In contrast, \name focuses on managed-language applications, and it identifies and addresses the challenges of extending existing managed runtimes to provide object-level CXL management for managed-language applications.

Hardware support for sub-page management has been explored to avoid software overheads in heterogeneous memory systems~\cite{kokolis2019pageseer,kotra2018chameleon,prodromou2017mempod,ryoo2017silc,sim2014transparent,wang2019supporting}. More recently, systems have proposed treating local memory as an L4 cache for CXL memory~\cite{lepers2023johnny,zhong20242lm,intelBreakingMemory}. However, these approaches require substantial hardware metadata, often in expensive SRAM, and lack the flexibility of software-based solutions.

Finally, prior work has explored fine-grained tiered-memory management inside the JVM in the context of NVM and far memory~\cite{wang2019panthera,akram2018write,wang2022memliner,nguyen2024polar}. While these systems demonstrate the opportunities of extending managed runtimes for tiered-memory management, they target media- or application-specific problems, such as NVM write endurance, far-memory GC cost, or Spark-specific placement. \name instead uses managed runtime mechanisms to address intrapage hotness skew in CXL memory across applications.

\section{Conclusion}
\label{sec:con}
We present \name, a system that demonstrates the potential of extending existing managed runtimes for object-level CXL memory management. \name identifies the key challenges in this direction and addresses them through accurate and lightweight object hotness tracking, a hybrid relocation approach, and careful hot-object compaction. Our JVM prototype shows that \name incurs low runtime overhead while effectively mitigating intrapage hotness skew, outperforming page-based CXL memory management systems.

\bibliographystyle{ACM-Reference-Format}
\bibliography{references}

\end{document}